\documentclass[aps,preprint,tightenlines,showpacs,nofootinbib]{revtex4}
\usepackage{epsfig}

\begin{document}

\title{The few body problem of the quark-model hadron structure}
\author{A. Valcarce}
\affiliation{Departamento de F\'\i sica Fundamental, Universidad de Salamanca, E-37008
Salamanca, Spain}
\author{J. Vijande}
\affiliation{Departamento de F\'{\i}sica At\'{o}mica, Molecular y Nuclear, Universidad de Valencia (UV)
and IFIC (UV-CSIC), Valencia, Spain.}
\author{H. Garcilazo}
\affiliation{Escuela Superior de F\'\i sica y Matem\'aticas,
Instituto Polit\'ecnico Nacional,
Edificio 9, 07738 M\'exico D.F., Mexico}
\author{T.F. Caram\'es}
\affiliation{Departamento de F\'\i sica Fundamental, Universidad de Salamanca, E-37008
Salamanca, Spain}
\date{\today}

\begin{abstract}
We discuss recent results of hadron spectroscopy and hadron-hadron
interaction within a quark model framework. New experimental data
could point to the important role played by higher order Fock
space components on low-energy observables. Our aim is to obtain a
coherent description of the low-energy hadron phenomenology to
constrain QCD phenomenological models and try to learn about
low-energy realizations of the theory. This long-range effort
is based in adapting few-body techniques to study many-quark
systems.
\end{abstract}

\maketitle

\section{Introduction}

The last decade has been especially challenging for hadron
structure. New experimental data have shaken our understanding
of QCD low-energy phenomenology. This earthquake had its 
epicenter on charmonium spectroscopy. 
Since Gell-Mann conjecture, most of the 
meson experimental data were classified as $q\overline{q}$ 
states according to $SU(N)$ irreducible representations.
Its simplicity had converted meson spectroscopy in an ideal system to learn about
the properties of QCD. 
Nevertheless a number of interesting issues remain still open
as for example the understanding of some $D$-meson data obtained on the 
B factories or the structure of the scalar mesons.
Some of the $D_s$ states can be hardly accommodated in a
pure $q\bar q$ description. They present a mass much lower than the
na\"ive prediction of constituent quark models. Besides,
the underlying structure of the scalar mesons 
is still not well established theoretically. 
These discrepancies gave rise to the hypothesis about possible contributions
of higher order Fock space components allowed by the Gell-Mann
classification. These ideas were simultaneously extended to the baryon sector,
enlightening some of the obscure remaining problems.

In this talk we give an overview of the old dream of getting a coherent
understanding of the low-energy hadron phenomenology~\cite{Val05}. We make emphasis
on two different aspects. Firstly, we will address the study of meson and
baryon spectroscopy in an enlarged Hilbert space considering four and five
quark components. This seems nowadays unavoidable to understand
the experimental data and it builds the bridge towards the description
in terms of hadronic degrees of freedom. Secondly, the same scheme used
to describe the hadron spectroscopy should be valid for describing
the low-energy hadron-hadron scattering. We review our recent efforts
to account for the strangeness $-1$ and $-2$ two- and three-baryon systems within the 
same scheme used for hadron spectroscopy and that has also demonstrated its 
validity for the nucleon-nucleon sector. 

In particular we will address three different problems. In the first
section we will discuss the charmonium spectroscopy above the threshold for
the production of $D$ mesons. The second section is devoted to study the effect 
of selected S-wave meson-baryon channels to 
match poor baryon mass predictions from quark models with data. 
Finally, the third section addresses the problem of the study
of two- and three-baryon systems with strangeness $-1$ and $-2$ by means
of a baryon-baryon interaction derived from the quark model used to study
the hadron spectroscopy. 

\section{Heavy mesons: Charmonium}

Charmonium has been used as the test bed to demonstrate the color
Fermi-Breit structure of quark atoms obeying the same principles
as ordinary atoms~\cite{Isg83}. Its nonrelativistic character 
($v/c \approx 0.2-0.3)$ gave rise to an amazing agreement between
experiment and simple quark potential model predictions
as $c\bar c$ states~\cite{Eic80}.
Close to the threshold for the production of charmed mesons 
quark-antiquark models required of an improved interaction ~\cite{Isg99}. 
The corrections introduced to the quark-antiquark spectra explained some deviations
observed experimentally \cite{Eic04}.

Since 2003, we have witnessed a growth of puzzling new 
charmonium mesons, like the well-established $X(3872)$ or the
not so well-established
$Y(4260)$, $Z(3930)$, $X(3940)$, $Y(3940)$, $X(4008)$, $X(4160)$, 
$X(4260)$, $Y(4350)$, and $Y(4660)$. 
In addition, the Belle Collaboration has reported the observation
of similar states with non-zero electric charge: the $Z(4430)$, the
$Z_1(4040)$ and the $Z_2(4240)$ that have not yet been
confirmed by other experiments~\cite{Ols09}.

These new states do not fit, in general, the simple predictions of the quark-antiquark
schemes and, moreover, they overpopulate the expected number of states in (simple) two-body theories.
This situation is not uncommon in particle physics. For example,
in the light scalar-isoscalar meson sector hadronic molecules offer a 
reasonable explanation of the experimental data~\cite{Jaf77}. Also,
the study of the $NN$ system above the pion production threshold 
required new degrees of freedom to be incorporated in the theory, either
as pions or as excited states of the nucleon, i.e., 
the $\Delta$~\cite{Pop87}. 
This discussion suggests that charmonium spectroscopy could be rather
simple below the threshold production of charmed mesons but  
much more complex above it. In particular, the coupling to 
the closest $(c\bar c)(n\bar n)$ system, referred to 
as {\it unquenching the na\"ive quark model}~\cite{Clo05},
could be an important spectroscopic ingredient. Besides,
hidden-charm four-quark states could 
explain the overpopulation of quark-antiquark theoretical states.

In an attempt to disentangle the role played by multiquark configurations 
in the charmonium spectroscopy we have solved the Lippmann-Schwinger equation 
for the scattering of two $D$ mesons looking for attractive channels 
that may contain a meson-meson molecule~\cite{Fer09}.
In order to account for all basis states we allow 
for the coupling to charmonium-light two-meson systems.
When we consider the system of two mesons $M_1$ and $\overline{M}_2$ ($M_i=D, D^*$)
in a relative $S-$state interacting through a potential $V$ that contains a
tensor force then, in general, there is a coupling to the
$M_1\overline{M}_2$ $D-$wave and the
Lippmann-Schwinger equation of the system is
\begin{equation}
t_{ji}^{\ell s\ell^{\prime \prime }s^{\prime \prime }}(p,p^{\prime \prime };E)
=V_{ji}^{\ell s\ell^{\prime \prime }s^{\prime \prime }}(p,p^{\prime \prime
})+\sum_{\ell^{\prime }s'}\int_{0}^{\infty }{p^{\prime }}%
^{2}dp^{\prime }
\, V_{ji}^{\ell s \ell^{\prime }s^{\prime}}
(p,p^{\prime })
{\frac{1}{E-{p^{\prime }}^{2}/2{\bf \mu }+i\epsilon }}%
t_{ji}^{\ell^{\prime }s^{\prime }\ell^{\prime \prime }s^{\prime \prime
}}(p^{\prime },p^{\prime \prime };E),  \label{eq1}
\end{equation}
where $t$ is the two-body amplitude, $j$, $i$, and $E$ are the
angular momentum, isospin and energy of the system, and $\ell s$,
$\ell^{\prime }s^{\prime }$, $\ell^{\prime \prime }s^{\prime \prime }$
are the initial, intermediate, and final orbital angular momentum
and spin; $p$
and $\mu $ are the relative momentum and reduced mass of the
two-body system, respectively.
In the case of a two $D$ meson system that can couple to a charmonium-light
two-meson state, for example when $D\overline{D}^*$ is coupled to
 $J/\Psi \omega$, the Lippmann-Schwinger equation for
$D\overline{D}^*$ scattering becomes
\begin{eqnarray}
t_{\alpha\beta;ji}^{\ell_\alpha s_\alpha \ell_\beta s_\beta}(p_\alpha,p_\beta;E) =
V_{\alpha\beta;ji}^{\ell_\alpha s_\alpha \ell_\beta s_\beta}(p_\alpha,p_\beta)&+&
\sum_{\gamma}\sum_{\ell_\gamma s_\gamma}
\int_0^\infty p_\gamma^2 dp_\gamma
V_{\alpha\gamma;ji}^{\ell_\alpha s_\alpha \ell_\gamma s_\gamma}
(p_\alpha,p_\gamma) \nonumber \\
&\times& \, G_\gamma(E;p_\gamma)
t_{\gamma\beta;ji}^{\ell_\gamma s_\gamma \ell_\beta s_\beta}
(p_\gamma,p_\beta;E),
\label{eq2}
\end{eqnarray}
with $\alpha, \beta, \gamma= D\overline{D}^*, J/\Psi \omega$.

\begin{figure}[b]
\caption{Fredholm determinant for the $J^{PC}(I)=1^{++}(0)$ $D\overline{D}^*$
system. Solid (dashed) line: results with (without) coupling to the $J/\Psi \omega$
channel.}
\vspace*{0.5cm}
\mbox{\epsfxsize=85mm\epsffile{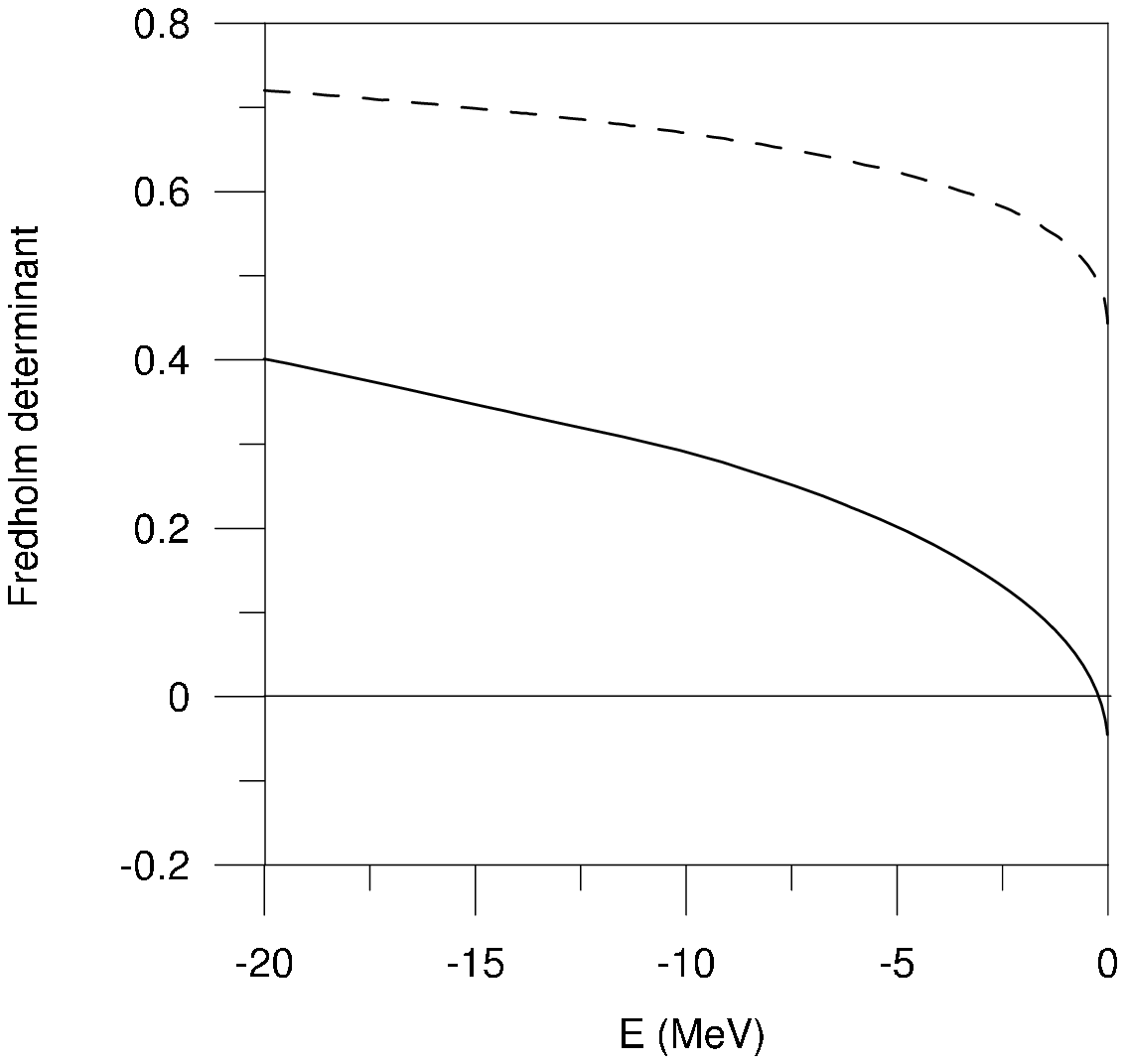}}
\label{f1}
\end{figure}

We have consistently used the same interacting Hamiltonian to study
the two- and four-quark systems to guarantee that thresholds
and possible bound states are eigenstates of the same Hamiltonian.
This study has been the first systematic analysis of four-quark
hidden-charm  meson-meson molecules. For the first
time a consistent study of all quantum numbers within the
same model was performed.
Our predictions robustly show that no deeply bound states can be expected for this system.
Only a few channels, see Table~\ref{t1}, can be expected to
present observable resonances or slightly bound states. Among them, we
have found that the $D\overline{D}^*$ system must show a bound state
slightly below threshold with quantum numbers $J^{PC}(I)=1^{++}(0)$,
that could correspond to the widely discussed $X(3872)$. 
\begin{table}[t]
\caption{Attractive channels for the two $D-$meson systems.}
{\begin{tabular}{@{}cccccc@{}} 
\toprule
System & $D\overline{D}$ & $D\overline{D}^*$ &\multicolumn{3}{c}{ $D^*\overline{D}^*$} \\ \colrule
$J^{PC}(I)$ &  $0^{++}(0)$ & $1^{++}(0)$ & $0^{++}(0)$ & $2^{++}(0)$ & $2^{++}(1)$ \\ \botrule
\end{tabular} \label{t1}}
\end{table}
This result is illustrated in Fig.~\ref{f1}. Out of the systems
made of a particle and its corresponding antiparticle,
$D\overline{D}$ and $D^*\overline{D}^*$, the $J^{PC}(I)=0^{++}(0)$
is attractive. It would be the only candidate to accommodate
a wide resonance for the $D\overline{D}$ system.
For the $D^*\overline{D}^*$ the attraction
is stronger and structures may be observed close and above the charmed
meson production threshold.
Also, we have shown that the $J^{PC}(I)=2^{++}(0,1)$ $D^*\overline{D}^*$
channels are attractive due to the coupling to the $J/\Psi \omega$ and
$J/\Psi \rho$ channels, respectively.

Among these exotic theoretical states, charged states have an
unique feature: by construction they cannot be accommodated into
the conventional $c\bar c$ spectrum. Two different experimental
findings show positive results on charge charmonium mesons.
The first one was a $\Psi(2S)\pi^+$ peak at about
4430 MeV/c$^2$ observed by Belle in the
$\bar B^0 \to \Psi (2S)\pi^+K^-$ decays~\cite{Cho08}. A second positive
observation was reported by Belle from the $B^0 \to \Xi_{c1} \pi^+K^-$ decay,
with two resonances in $\Xi_{c1} \pi^+$ at masses of about
4050 and 4250 MeV/c$^2$~\cite{Miz08}.
In the first case, BaBar has presented its own analysis~\cite{Aub09}
performing a detailed study of the acceptance and possible
reflections concluding that no significant signal exists on
the data. While the two experiments made different conclusion,
the data itself seem to be in a reasonable agreement except
for the lower available statistics of the BaBar experiment.
The states found in Ref.~\cite{Miz08} could correspond to the
$D^*\overline{D}^*$ $J^{PC}(I)=2^{++}(1)$ we
have predicted~\cite{Car10}.
Its confirmation would represent a unique tool in discriminating
among different theoretical models.

\section{Light baryons}

As we have seen above, there is nowadays compelling evidence of 
meson resonances containing more
than quark-antiquark $(q\overline{q})$ valence
components. One can guess a similar situation for some baryon resonances
that may contain more than three-quark $(3q)$ valence components.
In the baryon sector a
paradigmatic case is the $\Lambda (1405)$ which requires the consideration
of a $N\overline{K}$ component for its explanation. In the meson sector
scalars such as the f$_{0}$'s contain relevant meson-meson components. In
all these cases valence quark models based on\ $3q$ or $q\overline{q}$
components that describe correctly the bulk of spectral data fail
systematically to reproduce their masses. Actually we can interpret the
systematic failure in the description of a given resonance by valence quark
models as an indication of its anomalous nature in the sense of containing
more than the valence components. Here we use this interpretation as a
criteria to identify possible anomalies in the Light-quark Baryon Spectrum
(LqBS).
\begin{figure}[bp]
\caption{Mass predictions for the anomalies from Ref.~\protect\cite{Cap86}
(dashed lines) as compared to the experimental mass intervals (boxes). N.C.
means non-cataloged resonance. Meson-baryon thresholds are indicated by
solid lines.}
\vspace*{1cm}
\hspace*{2cm}\epsfig{file=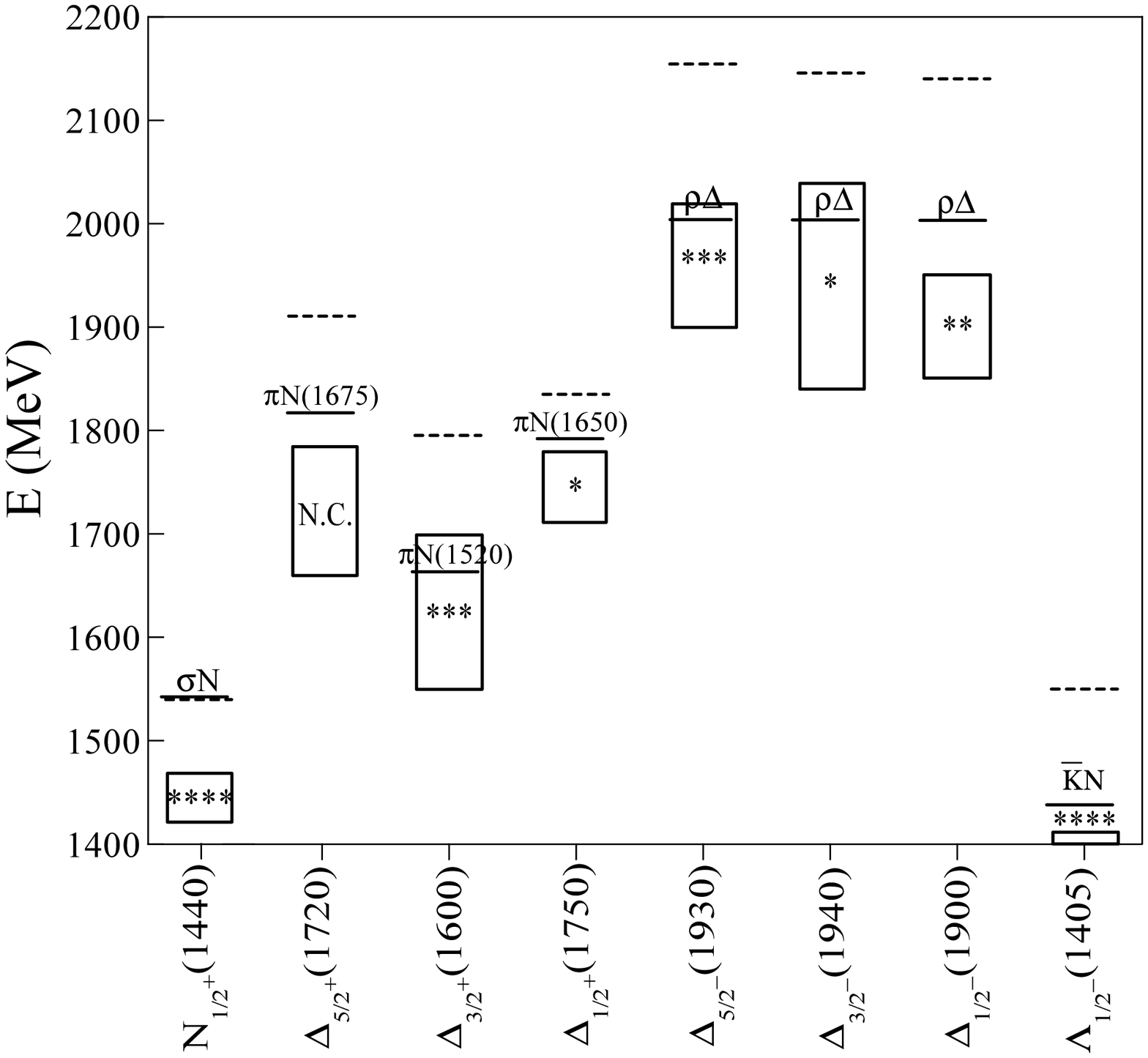,width=9cm}
\label{f2}
\end{figure}

\begin{figure}[tbp]
\caption{Predicted masses for the anomalies (dashed lines) as compared to
the experimental mass intervals (boxes). N.C. means non-cataloged resonance.}
\label{f3}
\vspace*{1cm}
\hspace*{2cm}\epsfig{file=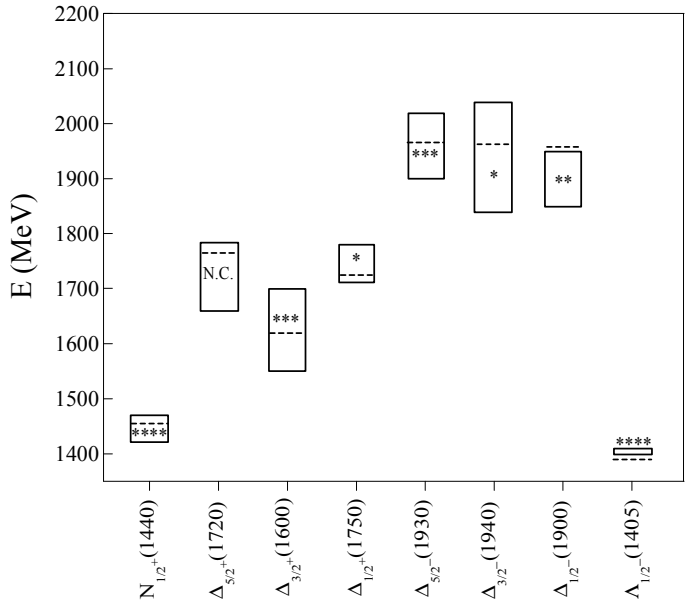,width=9cm}
\end{figure}

A comparative analysis of Constituent Quark Models (CQM) predictions for the
LqBS has been carried out in Ref.~\cite{GVV08}. This analysis makes clear
the presence of anomalies for which $3q$ predicted masses are significantly
higher than their Particle Data Group (PDG) averages~\cite{PDG08}, see
Fig.~\ref{f2}. Moreover all the anomalies have in common the existence of $S-$
wave meson-baryon thresholds with the same quantum numbers than the
anomalies (also drawn in Fig.~\ref{f2}) close above the PDG average
masses. This suggests a possible contribution of these meson-baryon ($4q1%
\overline{q})$ components to the anomalies to give a correct account of
their masses. To make a rough estimate of this contribution we shall
consider the coupling of the free meson-baryon $(mb)$ channels to the
corresponding $3q$ states. To make effective such a coupling we parametrize
the transition matrix element, $<mb\left\vert H\right\vert 3q>\equiv a,$ and
diagonalize the hamiltonian matrix by assuming $<3q\left\vert H\right\vert
3q>=M_{3q}$ being $M_{3q}$ the predicted $3q$ mass and 
$<mb\left\vert H\right\vert mb>\sim M_{m}+M_{b}$ being $M_{m}+M_{b}$ the mass of the
meson-baryon threshold. The results from the diagonalization corresponding
to the lowest energy states are shown in Fig.~\ref{f3}. Despite the
shortcomings of our toy model calculation the quantitative agreement with
data is spectacular (there is only one universal parameter, $a,$ assumed for
simplicity to be the same in all cases). Qualitatively our toy model implies
that the anomalies correspond mostly to meson-baryon states but with a
non-negligible three-quark probability that makes them stable against decay
into $m+b.$ This interpretation could need some refinement when going to a
more complete model where the meson-baryon interaction were also considered (%
$<mb\left| H\right| mb>\sim M_{m}+M_{b}).$ Then some anomalies
might appear as simply meson-baryon bound states. Keeping this in mind we
may conclude that the coupling to selected $mb$ channels acts as an
effective healing mechanism to match deficient CQM predictions with data.
Furthermore the implementation of these selected channels in the analyses of
data might add certainty to the existence of the anomalies and at the same
time help to reconcile competing and sometimes not compatible partial wave
analyses. Indeed the incorporation of some of our effective inelastic $mb$
channels in data analyses has been very relevant for the experimental
extraction of the corresponding anomalies: $\sigma N$ for $N_{P_{11}}(1440)$~%
\cite{Man92} and $\rho \Delta $ for $\Delta _{D_{35}}(1930)$~\cite%
{Man92,Cut80}.

\section{Strangeness $-1$ and $-2$ two and three-baryon systems}

In this section, we 
review recent studies of two and three-baryon systems with strangeness $-1$ and
$-2$ using two-body interactions derived from the quark model.

\subsection{The two-body interactions}

The baryon-baryon interactions are obtained from the chiral 
constituent quark model \cite{Val05,Gar05}. 
In order to derive the local $B_1B_2\to B_3B_4$ potentials from the
basic $qq$ interaction defined above we use a Born-Oppenheimer
approximation. Explicitly, the potential is calculated as follows,
\begin{equation}
V_{B_1B_2 (L \, S \, T) \rightarrow B_3B_4 (L^{\prime}\, S^{\prime}\, T)} (R) =
\xi_{L \,S \, T}^{L^{\prime}\, S^{\prime}\, T} (R) \, - \, \xi_{L \,S \,
T}^{L^{\prime}\, S^{\prime}\, T} (\infty) \, ,  \label{Poten1}
\end{equation}
\noindent where
\begin{equation}
\xi_{L \, S \, T}^{L^{\prime}\, S^{\prime}\, T} (R) \, = \, {\frac{{\left
\langle \Psi_{B_3B_4 }^{L^{\prime}\, S^{\prime}\, T} ({\vec R}) \mid
\sum_{i<j=1}^{6} V_{qq}({\vec r}_{ij}) \mid \Psi_{B_1B_2 }^{L \, S \, T} ({\vec R%
}) \right \rangle} }{{\sqrt{\left \langle \Psi_{B_3B_4 }^{L^{\prime}\,
S^{\prime}\, T} ({\vec R}) \mid \Psi_{B_3B_4 }^{L^{\prime}\, S^{\prime}\, T} ({%
\vec R}) \right \rangle} \sqrt{\left \langle \Psi_{B_1B_2}^{L \, S \, T} ({\vec %
R}) \mid \Psi_{B_1B_2}^{L \, S \, T} ({\vec R}) \right \rangle}}}} \, .
\label{Poten2}
\end{equation}
In the last expression the quark coordinates are integrated out keeping $R$
fixed, the resulting interaction being a function of the $B_i-B_j$ 
relative distance. The wave function 
$\Psi_{B_iB_j}^{L \, S \, T}({\vec R})$ for the two-baryon system
is discussed in detail in Ref. \cite{Val05}.

In Ref.~\cite{GFV07} we constructed 
different families of $S=-1$ interacting potentials,
by introducing small variations of the mass of the
effective scalar exchange potentials, 
that allow us to study the dependence of the results on the
strength of the spin-singlet and spin-triplet 
hyperon-nucleon interactions. 
These potentials are characterized by
the $\Lambda N$ scattering lengths $a_{i,s}$
and they reproduce
the cross sections near threshold of the five
hyperon-nucleon processes for which data are 
available (see Ref.~\cite{GFV07}). Let us analyze
the three-body systems with strangeness $-1$.

\subsection{The $\Lambda NN$ system}

The channels $(I,J)$ = (0,1/2) and (0,3/2) are the most attractive ones of
the $\Lambda NN$ system. In particular, the channel (0,1/2) has the only 
bound state of this system, the hypertriton. We give in Table~\ref{t2}
the results of the models constructed in Ref.~\cite{GFV07} for the
two $\Lambda d$ scattering lengths and the hypertriton binding energy.
\begin{table}[tbh!]
\caption{$\Lambda d$ scattering lengths, $A_{0,3/2}$ and $A_{0,1/2}$ (in fm),
and hypertriton binding energy,
$B_{0,1/2}$ (in MeV), for several hyperon-nucleon interactions
characterized by $\Lambda N$ scattering lengths
$a_{1/2,0}$ and $a_{1/2,1}$ (in fm).}
{\begin{tabular}{@{}ccccc@{}} 
\toprule
$a_{1/2,0}$  &  $a_{1/2,1}$ &
 $A_{0,3/2}$ & $A_{0,1/2}$  & $B_{0,1/2}$ \\
\colrule
2.48 & 1.41 & 31.9 (66.3)        & $-$16.0 ($-$20.0) & 0.129 (0.089) \\
2.48 & 1.65 & $-$72.8 (198.2)    & $-$13.8 ($-$17.2) & 0.178 (0.124) \\
2.48 & 1.72 & $-$40.8 ($-$179.8) & $-$13.3 ($-$16.6) & 0.192 (0.134) \\
2.48 & 1.79 & $-$28.5 ($-$62.7)  & $-$12.9 ($-$16.0) & 0.207 (0.145) \\
2.48 & 1.87 & $-$22.0 ($-$38.2)  & $-$12.5 ($-$15.4) & 0.223 (0.156) \\
2.48 & 1.95 & $-$17.9 ($-$27.6)  & $-$12.1 ($-$14.9) & 0.239 (0.168) \\
2.31 & 1.65 & $-$76.0 (198.2)    & $-$17.1 ($-$22.4) & 0.113 (0.070) \\
2.55 & 1.65 & $-$73.6 (198.2)    & $-$13.6 ($-$16.8) & 0.185 (0.130) \\
2.74 & 1.65 & $-$72.1 (198.2)    & $-$12.0 ($-$14.4) & 0.244 (0.182) \\ \botrule
\end{tabular} \label{t2}}
\end{table}
We compare with the results, in parentheses, obtained in
Ref.~\cite{GFV07} including only the three-body $S$ wave configurations.
As a consequence of considering the $D$ waves, the hypertriton binding energy
increases by about 50$-$60 keV~\cite{Fuj07}, 
while the $A_{0,1/2}$ scattering length
decreases by about 3$-$5 fm. The largest changes occur in the $A_{0,3/2}$
scattering length where both positive and negative values appeared which
means, in the case of the negative values, that a bound state is
generated in the $(I,J)=(0,3/2)$ channel. Since this channel depends
mainly on the spin-triplet hyperon-nucleon interaction and experimentally
there is no evidence whatsoever for the existence of a $(I,J)=(0,3/2)$ 
bound state
one can use the results of this channel to set limits on the value of
the hyperon-nucleon spin-triplet scattering length $a_{1/2,1}$. 
As one can guess from Table~\ref{t2}, the
three-body channel $(I,J)=(0,3/2)$ becomes bound
if $a_{1/2,1} > 1.58$ fm (see Ref.~\cite{OND07}.
Moreover, we found in Ref.~\cite{GFV07} that the fit 
of the hyperon-nucleon cross sections is worsened 
for those cases where the spin-triplet $\Lambda N$
scattering length is smaller than 1.41 fm,
so that we conclude that $1.41\le a_{1/2,1} \le 1.58$ fm.
To set some limits to the
hyperon-nucleon spin-singlet scattering length,
we have calculated in Table~\ref{t3} the
hypertriton binding energy using for the hyperon-nucleon spin-triplet
scattering length the allowed values $1.41\le a_{1/2,1}\le 1.58$ fm
and for the spin-singlet scattering length 
$2.33\le a_{1/2,0}\le 2.48$ fm which leads to results for the 
hypertriton binding energy within the experimental error bars
$B_{0,1/2}=0.13\pm 0.05$ MeV.
\begin{table}[tp!]
\caption{Hypertriton binding energy
(in MeV) for several hyperon-nucleon interactions
characterized by $\Lambda N$ scattering lengths
$a_{1/2,0}$ and $a_{1/2,1}$ (in fm) which are within the
experimental error bars $B_{0,1/2}=0.130\pm 0.050$ MeV.}
{\begin{tabular}{@{}ccccc@{}} 
\toprule
& $a_{1/2,1}=1.41$  &  $a_{1/2,1}=1.46$ 
& $a_{1/2,1}=1.52$  &  $a_{1/2,1}=1.58$ \\
\colrule
$a_{1/2,0}=2.33$ & 0.080 & 0.087   & 0.096 & 0.106 \\
$a_{1/2,0}=2.39$ & 0.094 & 0.102   & 0.112 & 0.122 \\
$a_{1/2,0}=2.48$ & 0.129 & 0.140   & 0.152 & 0.164 \\ \botrule
\end{tabular}\label{t3}}
\end{table}
\begin{table}[bp!]
\caption{$\Sigma d$ scattering lengths,
$A_{1,3/2}^\prime$ and $A_{1,1/2}^\prime$ (in fm),
and position of the quasibound state $B_{1,1/2}^\prime$
(in MeV) for several hyperon-nucleon interactions
characterized by $\Lambda N$ scattering lengths
$a_{1/2,0}$ and $a_{1/2,1}$ (in fm).}
{\begin{tabular}{@{}ccccc@{}} 
\toprule
$a_{1/2,0}$  &  $a_{1/2,1}$ &
 $A_{1,3/2}^\prime$ & $A_{1,1/2}^\prime$ & $B_{1,1/2}^\prime$ \\ 
\colrule     
2.48 & 1.41 & 0.14$+\, i\, $0.24 (0.20$\, +i\, $0.26) & 19.82$+\, i\, $16.94 (19.28$+\, i\, $25.37)       &2.92$-\, i\, $2.17 \\
2.48 & 1.65 & 0.28$+\, i\, $0.27 (0.36$\, +i\, $0.29) & 12.08$+\, i\, $38.98 ($-$1.55$+\, i\, $42.31)     &2.84$-\, i\, $2.14 \\
2.48 & 1.72 & 0.32$+\, i\, $0.28 (0.40$\, +i\, $0.30) & 2.92$+\, i\, $43.20 ($-$10.47$+\, i\, $40.25)     &2.82$-\, i\, $2.11 \\
2.48 & 1.79 & 0.36$+\, i\, $0.29 (0.44$\, +i\, $0.31) &$-$8.00$+\, i\, $42.58 ($-$17.33$+\, i\, $35.01)   &2.79$-\, i\, $2.10 \\
2.48 & 1.87 & 0.40$+\, i\, $0.30 (0.49$\, +i\, $0.33) &$-$16.90$+\, i\, $37.08 ($-$21.16$+\, i\, $28.54)  &2.77$-\, i\, $2.09 \\
2.48 & 1.95 & 0.45$+\, i\, $0.31 (0.54$\, +i\, $0.34) &$-$21.73$+\, i\, $29.48 ($-$22.44$+\, i\, $22.44)  &2.75$-\, i\, $2.08 \\
2.31 & 1.65 & 0.28$+\, i\, $0.27 (0.36$\, +i\, $0.29) & 19.01$+\, i\, $23.21 (14.95$+\, i\, $31.61)       &2.88$-\, i\, $2.14\\
2.55 & 1.65 & 0.28$+\, i\, $0.27 (0.36$\, +i\, $0.29) & $-$12.81$+\, i\, $43.49 ($-$21.04$+\, i\, $33.19) &2.79$-\, i\, $2.11\\
2.74 & 1.65 & 0.28$+\, i\, $0.27 (0.36$\, +i\, $0.29) & $-$26.01$+\, i\, $17.95 ($-$23.29$+\, i\, $13.32) &2.73$-\, i\, $2.09\\ \botrule
\end{tabular} \label{t4}}
\end{table}

With regard to the isospin 1 channels $(I,J)=(1,1/2)$ and $(1,3/2)$,
the $(1,1/2)$ channel
is attractive but not enough to produce a bound state while the $(1,3/2)$
channel is repulsive. 

\subsection{The $\Sigma NN$ system}
We show in Table~\ref{t4} the $\Sigma d$ scattering lengths 
$A_{1,3/2}^\prime$ and $A_{1,1/2}^\prime$.
The $\Sigma d$ scattering lengths are 
complex since the inelastic $\Lambda NN$ channels are always open.
The scattering length $A_{1,3/2}^\prime$ depends mainly on the
spin-triplet hyperon-nucleon channels and both its real and
imaginary parts increase when the spin-triplet hyperon-nucleon
scattering length increases. The effect of 
the three-body $D$ waves is to lower 
the real part by about 20 \% and the imaginary part by about 10 \%. 
The scattering length $A_{1,1/2}^\prime$ shows large variations 
between the results with and without 
three-body $D$ waves but this is due, as
we will see next, to the fact that there is a pole very near threshold,
a situation quite similar to that of the $A_{0,3/2}$ $\Lambda d$
scattering length discussed in the previous subsection. 
\begin{figure}[tp!]
\caption{Real and imaginary parts of the $\Sigma d$ scattering
length $A_{1,1/2}^\prime$
as a function of the $\Lambda N$ $a_{1/2,1}$ scattering length.}
\mbox{\epsfxsize=110mm\epsffile{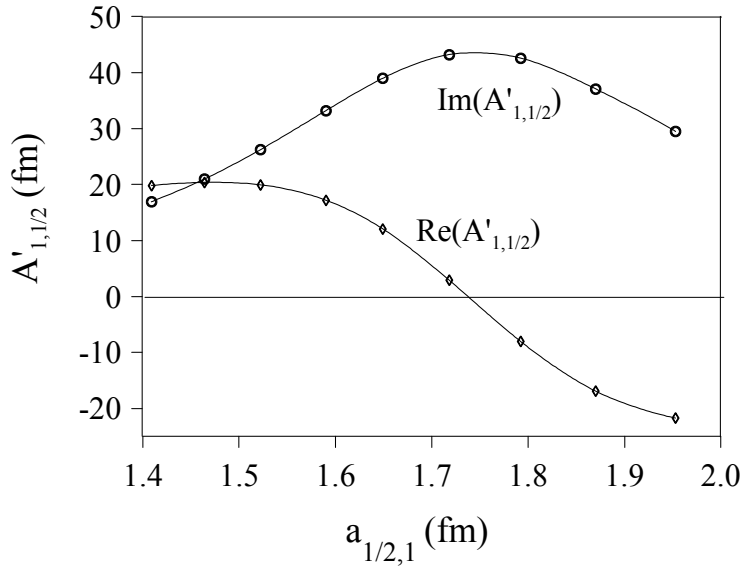}}
\vspace*{-8cm}
\label{f4}
\end{figure}

We plot in Fig.~\ref{f4} the real and imaginary parts of the $\Sigma d$
scattering length $A_{1,1/2}^\prime$ as functions of the spin-triplet
$\Lambda N$ scattering length $a_{1/2,1}$, since by increasing 
$a_{1/2,1}$ one is increasing the amount of attraction that is
present in the three-body channel. As one can see, Re$(A_{1,1/2}^\prime)$
changes sign from positive to negative while at the same time
Im$(A_{1,1/2}^\prime)$ has a maximum. These two features are the typical
ones that signal that the channel has a quasibound state \cite{DELOF}.
The position of this pole in the complex plane, which is given in the last
column of Table~\ref{t4}, changes
very little with the model used and it lies at around
2.8$\, -i\, $2.1 MeV. 

\subsection{Strangeness -2 two-baryon systems}

The knowledge of the strangeness $S=-2$ two-baryon interactions 
has become an important issue for theoretical and experimental 
studies of the strangeness nuclear physics. Moreover, this is
an important piece of a more fundamental problem, the description
of the interaction of the different members of the baryon octet
in a unified way. The $\Xi N -\Lambda \Lambda$ interaction accounts
for the existence of doubly strange hypernuclei, which is a
gateway to strange hadronic matter. Strangeness $-2$ baryon-baryon 
interactions also account for a possible six-quark $H$-dibaryon, 
which has yet to be experimentally observed.

There has been a steady progress towards the $S=-2$ baryon-baryon interaction.
The $\Lambda N$ interaction is pretty much understood based on the 
experimental data of $\Lambda$ hypernuclei. There is also 
some progress made on the $\Sigma N$ interaction. However,
the experimental knowledge on the $\Xi N$ and hyperon-hyperon ($YY$) interactions
is quite poor. The only information available came from doubly
strange hypernuclei, suggesting that the $^1S_0$ $\Lambda\Lambda$
interaction should be moderately attractive. An upper limit of
$B_{\Lambda \Lambda}=7.25\pm0.19$ MeV has been deduced for a 
hypothetical $H$-dibaryon (a lower limit of $2223.7$ MeV/$c^2$
for its mass) from
the so-called Nagara event~\cite{Nag08} at a 90\% confidence level.
The KEK-E176/E373 hybrid emulsion experiments observed other 
events corresponding to double$-\Lambda$ hypernuclei. Among
them, the Demachi-Yanagi~\cite{Nak09,Ich01}, identified as $^{10}_{\Lambda \Lambda}$Be
drove a value of $B_{\Lambda\Lambda}=11.90 \pm 0.13$~\cite{Hiy10}. A new
observed double-$\Lambda$ event has been recently reported by the
KEK-E373 experiment, the Hida event~\cite{Ich01},
although with uncertainties on its nature it has been
interpreted as an observation of the ground state 
of the $^{11}_{\Lambda \Lambda}$Be~\cite{Hiy10}.
Very recently, doubly strange baryon-baryon scattering data at
low energies were deduced for the first time, obtaining some
experimental data and upper limits for different elastic and inelastic 
cross sections: $\Xi^-p \to \Xi^- p$ and $\Xi^-p \to \Lambda \Lambda$~\cite{Ahn06}.
Recent results of the KEK-PS E522 experiment indicate the possibility
of a $H$-dibaryon as a resonance state
with a mass range between the $\Lambda\Lambda$ and
$N\Xi$ thresholds~\cite{Yoo07}. 

In the planned experiments at J-PARC~\cite{Nag00}, dozens of 
emulsions events for double$-\Lambda$ hypernuclei will be produced. 
The $(K^-,K^+)$ reaction is one of the most promising ways of studying
doubly strange systems. $\Lambda\Lambda$ hypernuclei can be produced
through the reaction $K^- p \to K^+ \Xi^-$ followed by $\Xi^- p \to \Lambda\Lambda$.
It is therefore compulsory having theoretical predictions concerning
the $\Xi N - \Lambda\Lambda$ coupling~\cite{Har10} to guide the
experimental way and, in a symbiotic manner, to use the new experimental data
as a feedback of our theoretical knowledge. 

\begin{figure}[tbp!]
\caption{$\Xi^- p$ elastic cross section, in mb, as a function of
the laboratory $\Xi$ momentum, in GeV/c. The two experimental data are taken
from Ref.~\cite{Tam01}, black circle, and Ref.~\cite{Yam01}, black square 
(both are in-medium experimental data). The horizontal 
line indicates an upper limit for the cross section extracted in Ref.~\cite{Ahn06}
with a large uncertainty in the momentum.}
\mbox{\epsfxsize=110mm\epsffile{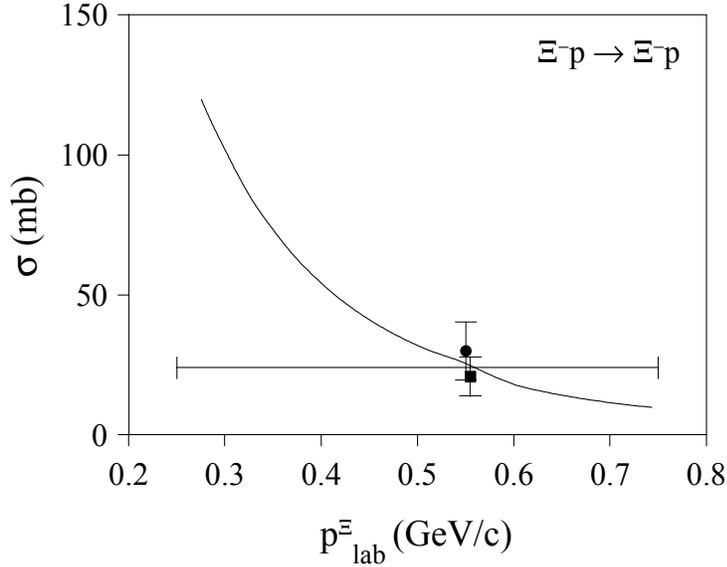}}
\vspace*{-8cm}
\label{f5}
\end{figure}
\begin{figure}[bp!]
\caption{$\Xi^- p \to \Lambda\Lambda$ cross section, in mb, as a function of
the laboratory $\Xi$ momentum, in GeV/c. The experimental data is taken
from Ref.~\cite{Ahn06}.}
\mbox{\epsfxsize=110mm\epsffile{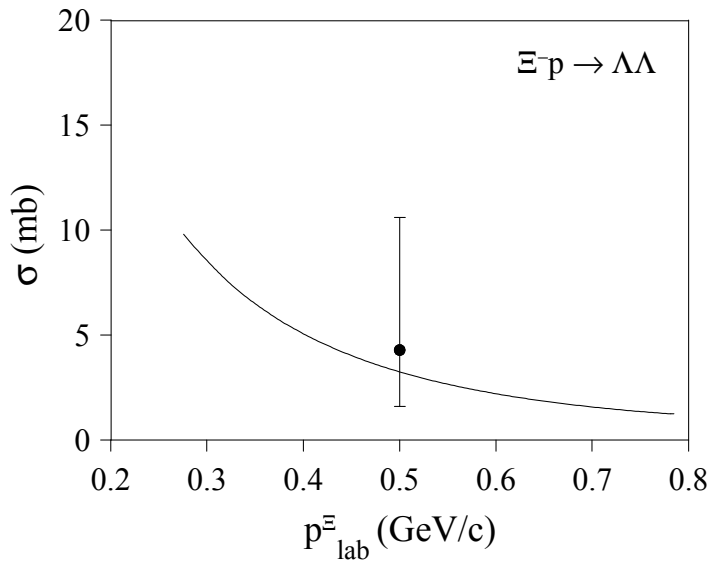}}
\vspace*{-8cm}
\label{f6}
\end{figure}
\begin{figure}[bp!]
\caption{$\Xi^- p \to \Xi^0 n$ cross section, in mb, as a function of
the laboratory $\Xi$ momentum, in GeV/c. The experimental data
are taken from Ref.~\cite{Ahn06} as explained in the text.}
\mbox{\epsfxsize=110mm\epsffile{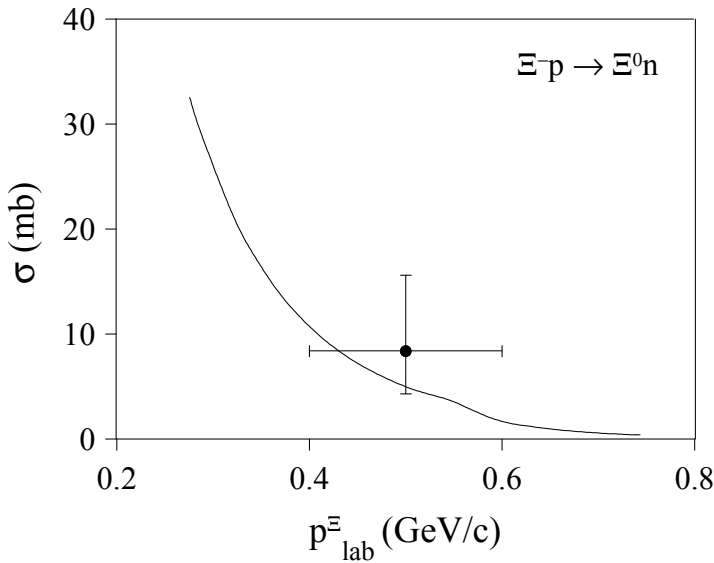}}
\vspace*{-8cm}
\label{f7}
\end{figure}

In this section, we derive the strangeness
$-2$ baryon-baryon interactions: $\Lambda \Lambda$, $\Lambda\Sigma$,
$\Sigma \Sigma$ and $N\Xi$.
We use these two-body interactions
to calculate two-body elastic
and inelastic scattering cross sections~\cite{PLB10}
and we compare to experimental data and other theoretical models.
We  also calculate the two-body
scattering lengths of the different spin-isospin channels
to compare with other theoretical models.
We show in Fig.~\ref{f5} the $\Xi^- p$ elastic cross section compared
to the in-medium experimental $\Xi^- p$ cross section around $p^\Xi_{\rm lab}=$
550 MeV/c, where $\sigma_{\Xi^- p}=$ 30 $\pm$ 6.7 $^{+3.7}_{-3.6}$ mb~\cite{Tam01}.
Another analysis using the eikonal approximation gives
$\sigma_{\Xi^- p}=$ 20.9 $\pm$ 4.5 $^{+2.5}_{-2.4}$ mb~\cite{Yam01}. A more recent
experimental analysis~\cite{Ahn06} for the low energy $\Xi^- p$ elastic
and $\Xi^- p \to \Lambda \Lambda$ total cross sections in the range
0.2 GeV/c to 0.8 GeV/c shows that the former is less than 24 mb at 90\%
confidence level and the latter of the order of several mb, respectively.
In Fig.~\ref{f6} we present the inelastic $\Xi^- p \to \Lambda\Lambda$
cross section. It has been recently estimated at a laboratory momentum
of $p^\Xi_{\rm lab}=$ 500 MeV/$c$, see Ref.~\cite{Ahn06}, assuming a quasifree 
scattering process for the reaction $^{12}{\rm C}(\Xi^-,\Lambda\Lambda)X$ 
obtaining a total cross section
$\sigma(\Xi^- p \to \Lambda \Lambda)=4.3^{+6.3}_{-2.7}$ mb. 
The upper limit of the cross section was derived as 12 mb at 90\% confidence level.
Fig.~\ref{f7} shows the total inelastic cross section $\Xi^- p \to \Xi^0 n$.
Combining the results of Refs.~\cite{Ahn06} and ~\cite{Ahn98}, one obtains
$\sigma(\Xi^- p \to \Xi^0 n) \sim 10$ mb. A recent measurement of a 
quasifree $p(K^-,K^+)\Xi^-$ reaction in emulsion plates yielded
$12.7^{+3.5}_{-3.1}$ mb for the total inelastic cross section in
the momentum range $0.4-0.6$ GeV/$c$~\cite{Aok98}, consistent with
the results of Ref.~\cite{Ahn98}. The experimental results for
$\Xi^- p$ inelastic scattering of Refs.~\cite{Ahn98,Aok98} involve
both $\Xi^- p \to \Lambda \Lambda$ and $\Xi^- p \to \Xi^ 0 n$~\cite{Ahn06},
what combined with the value for $\Xi^- p \to \Lambda\Lambda$ of Ref.~\cite{Ahn06}
allows to obtain an estimate of the inelastic cross section $\Xi^- p \to \Xi^0 n$.

As can be seen our results agree with the
experimental data for the elastic and inelastic $\Xi N$
cross sections. The small bumps in the cross sections
correspond to the opening of inelastic channels.
We would like to emphasize the agreement of our
results with the $\Xi^- p \to \Lambda \Lambda$ conversion cross section.
This reaction is of particular importance in assessing the stability
of $\Xi^-$ quasi-particle states in nuclei. 
Our results are close to the estimations of the Nijmegen-D model~\cite{NijD}.
Ref.~\cite{Fuj07} predicts $\sigma (\Xi^- p \to \Xi^0 n)\sim$
15 mb at $p^\Xi_{\rm lab}=$ 0.5 GeV/$c$. Ref.~\cite{Ahn98}
reported 14 mb for the inelastic scattering involving
both $\Xi^- p \to \Lambda \Lambda$ and $\Xi^- p \to \Xi^0 n$.
We found a smaller 
value of approximately 6 mb for both inelastic channels,
in close agreement to experiment.

The scattering lengths for the different spin-isospin
channels are given in Table~\ref{t5}. 
\begin{table}[tbp!]
\caption{Two-body singlet and triplet scattering lengths, in fm, for different
models as compared to our results. The results between squared brackets indicate the
lower and upper limit for different parametrizations used in that reference.}
{\begin{tabular}{@{}cccccc@{}} 
\toprule
Model & Ref.~\cite{Sto99} & Ref.~\cite{Rij06} & Ref.~\cite{Pol07} & Ref.~\cite{Fuj07} & Ours~\cite{PLB10} \\
\colrule     
$a_{^1S_0}^{\Lambda\Lambda}$  & $[-0.27,-0.35]$ & $[-1.555,-3.804]$ & $[-1.52,-1.67]$ & $-0.821$ & $-2.54$ \\
$a_{^1S_0}^{\Xi^0 p}$         & $[0.40,0.46]$   & $[0.144,0.491]$  & $[0.13,0.21]$  & $0.324 $ & $-3.32$ \\
$a_{^3S_1}^{\Xi^0 p}$         & $[-0.030,0.050]$  & $-$   & $[0.0,0.03]$   & $-0.207$ & $18.69$ \\
$a_{^1S_0}^{\Sigma^+\Sigma^+}$& $[6.98,10.32]$  & $-$           & $[-6.23,-9.27]$& $-85.3$      & $0.523$ \\ \hline
$a_{^1S_0}^{\Xi N (I=0)}$     &$   -       $   & $     -        $  & $   -       $  & $  -  $ & $-1.20\, + \, i \, 0.75$ \\
$a_{^3S_1}^{\Xi N (I=0)}$         &$   -        $  & $[-1.672,122.5]$   & $   -      $   & $  -   $ & $0.28$ \\
$a_{^1S_0}^{\Lambda\Sigma}$   &$   -       $   & $     -        $  & $   -       $  & $  - $ & $0.908\, + \, i \, 1.319$ \\
$a_{^3S_1}^{\Lambda\Sigma}$   &$   -   $   & $     -   $  & $   -       $  & $  - $ & $ - $ $3.116\, + \, i \, 0.393$ \\
$a_{^3S_1}^{\Sigma\Sigma (I=1)}$   &$   -   $   & $     -   $  & $   -       $  & $  - $ & $ - $ $1.347\, + \, i \, 1.801$ \\
$a_{^1S_0}^{\Sigma\Sigma (I=0)}$   &$   -   $   & $     -   $  & $   -       $  & $  - $ & $ - $ $0.039\, + \, i \, 0.517$ \\ \botrule
\end{tabular} \label{t5}}
\end{table}
These parameters are complex 
for the $N\Xi$ $(i,j)=(0,0)$ since the inelastic $\Lambda \Lambda$
channel is open, for the $\Lambda \Sigma$ isospin 1 channel due
to the opening of the $N\Xi$ channel, and for the 
$\Sigma\Sigma$ $i=$0 and 1 since the $\Lambda\Lambda$
and $N\Xi$ channels are open in the first case, and the $N\Xi$ and $\Lambda\Sigma$
are open in the second.  
There is no direct comparison between our results and those of
Ref.~\cite{Fuj07}. Although both are quark-model based results,
Ref.~\cite{Fuj07} used an old-fashioned quark-model interacting potential.
For example, they use a huge
strong coupling constant, $\alpha_S=1.9759$, and they consider the contribution of
vector mesons what could give rise to double counting problems~\cite{Yaz90}. 
As explicitly written in Ref.~\cite{Fuj07}, the parameters
are effective in their approach and has very little to do with QCD.
As mentioned above, since the observation
of the Nagara event~\cite{Nag08} it is generally accepted
that the $\Lambda\Lambda$ interaction is only moderately attractive.
Our result for the $^1S_0$ $\Lambda\Lambda$ scattering length is
compatible with such event. A rough estimate of $B_{\Lambda\Lambda}(
B_{\Lambda\Lambda}\approx (\hbar c)^2 / 2 \mu_{\Lambda\Lambda} a^{\Lambda\Lambda}_{^1S_0}$)
drives a value of 5.41 MeV, below the upper limit extracted from the Nagara
event, $B_{\Lambda\Lambda}=7.25\pm0.19$ MeV. Moreover, although from
the $\Lambda\Lambda$ scattering length alone one cannot draw any conclusion
on the magnitude of the two-$\Lambda$ separation energy, recent 
estimates~\cite{Rij06b} have reproduced the two-$\Lambda$ separation
energy, defined as $\Delta B_{\Lambda\Lambda}=B_{\Lambda\Lambda}(^6_{\Lambda\Lambda}{\rm He})
-2B_{\Lambda}(^5_\Lambda {\rm He})$, with scattering lengths of $-1.32$ fm.

\section{Summary}
\label{Sum}

To summarize, we have reviewed some recent results of hadron spectroscopy
and hadron-hadron interaction in terms of the constituent quark model.
The main goal of our presentation was to highlight the complementarity
of a simultaneous study of both problems in order to constrain low-energy
realizations of the underlying theory, QCD. We have seen how the enlargement
of the Hilbert space when increasing energy, that was seen to be necessary
to describe the nucleon-nucleon phenomenology above the pion production
threshold, seems to be necessary to understand current problems of hadron
spectroscopy. We have also tried to emphasize that spectroscopy and interaction
can and must be understood within the same scheme when dealing with quark
models. Any other alternative becomes irrelevant from the point of view
of learning about properties of QCD.

In our analysis we have shown the first systematic study of four-quark
hidden-charm states as meson-meson molecules. For the first
time a consistent study of all quantum numbers within the
same model has been done. 
Our predictions robustly show that no deeply bound states can be expected 
for charmonium.
Only a few channels can be expected to 
present observable resonances or slightly bound states. Among them, we
have found that the $D\overline{D}^*$ system must show a bound state
slightly below the threshold for charmed mesons production
with quantum numbers $J^{PC}(I)=1^{++}(0)$,
that could correspond to the widely discussed $X(3872)$. Out of the systems 
made of a particle and its corresponding antiparticle,
$D\overline{D}$ and $D^*\overline{D}^*$, the $J^{PC}(I)=0^{++}(0)$
is attractive. 
For the $D^*\overline{D}^*$ the attraction
is stronger and structures may be observed close and above the charmed
meson production threshold.
Also, we have shown that the $J^{PC}(I)=2^{++}(1)$ $D^*\overline{D}^*$
channel is attractive and may then represent a charge state contributing
to charmonium spestroscopy. 
Due to heavy quark symmetry, the replacement of the charm
quarks by bottom quarks decreases the kinetic energy without significantly
changing the potential energy. In consequence, four-quark bottomonium mesons
must also exist and have larger binding energies. 

We have seen how the $4q1\overline{q}$ components, in the form of $S$
wave meson-baryon channels which we identify, play an essential role in the
description of the anomalies, say baryon resonances very significantly
overpredicted by three-quark models based on two-body interactions. As a
matter of fact by considering a simplified description of the anomalies as
systems composed of a free meson-baryon channel interacting with a
three-quark confined component we have shown they could correspond mostly to
meson-baryon states but with a non-negligible $3q$ state probability which
makes their masses to be below the meson-baryon threshold. The remarkable
agreement of our results with data in all cases takes us to refine our
definition and propose the dominance of meson-baryon components as the
signature of an anomaly.
Though it is probable that these results may vary quantitatively when a more
complete dynamical coupled-channel calculation is carried out we think it is
reasonable not to expect major qualitative changes. 

We also presented results for the two and three-baryon systems with 
strangeness $-1$. We have solved the Faddeev equations for the $\Lambda NN$ and
$\Sigma NN$ systems using the hyperon-nucleon and nucleon-nucleon
interactions derived from a chiral constituent quark model with
full inclusion of the $\Lambda \leftrightarrow \Sigma$ conversion
and taking into account all three-body configurations
with $S$ and $D$ wave components.
In the case of the $\Lambda NN$ system the inclusion of the 
three-body $D$ wave components increases the attraction, reducing
the upper limit of the $a_{1/2,1}$ $\Lambda N$ scattering length
if the $(I,J)=(0,3/2)$ $\Lambda NN$ bound state does not exist.
This state shows a somewhat larger sensitivity than the
hypertriton to the three-body $D$ waves. Our calculation
including the three-body $D$ wave configurations of 
all relevant observables of two- and three-baryon 
systems with strangeness $-1$, permits to constrain the 
$\Lambda N$ scattering lengths to:
$1.41 \le a_{1/2,1} \le 1.58$ fm and
$2.33 \le a_{1/2,0} \le 2.48$ fm.
In the case of the $\Sigma NN$ system there exists a narrow quasibound
state near threshold in the $(I,J)=(1,1/2)$ channel. 

We have presented results for the doubly strange
$N \Xi $ and $Y_1Y_2$ interactions ($Y_i=\Sigma, \Lambda$).
We showed that the constituent quark model predictions are consistent with the recently
obtained doubly strange elastic and inelastic 
scattering cross sections. In particular, our results are compatible with the
$\Xi^- p \to \Lambda \Lambda$ conversion cross section,
important in assessing the stability
of $\Xi^-$ quasi-particle states in nuclei.

It is expected that in the coming years better-quality data on all sectors
discussed in this talk  will become available and our results can be used
to analyze these 
upcoming data in a model-independent way. 

\section{acknowledgments}
This work has been partially funded by the Spanish Ministerio de
Educaci\'on y Ciencia and EU FEDER under Contracts No. FPA2007-65748
and FPA2010-21750, by Junta de Castilla y Le\'{o}n under Contract No. GR12, 
and by the Spanish Consolider-Ingenio 2010 Program CPAN (CSD2007-00042),

\end{document}